\documentclass{aastex631}
\accepted{February 28, 2024}

\submitjournal{ApJ}

\begin{document}

\title{Gamma-Ray Burst Pulses and Lateral Jet Motion}

\correspondingauthor{Jon Hakkila}
\email{jh0271@uah.edu}

\author {Jon Hakkila}
\affiliation{University of Alabama in Huntsville,
Department of Physics and Astronomy,
Huntsville, AL 35899, USA}

\author{Geoffrey N. Pendleton}
\affiliation{DeciBel Research,
Huntsville, AL 35806, USA}

\author{Robert D. Preece}
\affiliation{University of Alabama in Huntsville,
Department of Space Sciences,
Huntsville, AL 35899, USA}

\author{Timothy W. Giblin}
\affiliation{511 Space Innovation LLC, Colorado Springs, CO 80919, USA}



\begin{abstract}

We propose that gamma-ray burst pulses are produced when highly-relativistic jets sweep across an observer's line-of-sight.
We hypothesize that axisymmetric jet profiles, coupled with special relativistic effects, produce the time-reversed properties of gamma-ray burst pulses. Curvature resulting from rapid jet expansion is responsible for much of the observed pulse asymmetry and hard-to-soft evolution. The relative obliqueness with which the jet crosses the line-of-sight explains the known GRB pulse morphological types. We explore two scenarios: one in which a rigid/semi-rigid jet moves laterally, and the other in which a ballistic jet sprays material from a laterally-moving nozzle. The ballistic jet model is favored based upon its consistency with standard emission mechanisms.

\end{abstract}



\section{Introduction} \label{sec:intro}

Despite decades of study, the mechanisms responsible for producing gamma-ray burst (GRB) prompt emission remain elusive. One reason for this is the inability of theoretical models to provide consistent explanations of GRB light curves characteristics. In recent years, improved data analysis techniques have allowed greater insights into GRB light curve behaviors. Some of the newly-identified behaviors, not found in other types of high energy transients, place peculiar constraints on GRB jet structure and thus also on GRB emission mechanisms.

\subsection{Pulses} \label{sec:pulses}

{\em Pulses} are the basic units of GRB prompt emission ({\em e.g.,} \cite{1996ApJ...459..393N, 2018ApJ...863...77H}). FRED (Fast Rise Exponential Decay) pulses were first recognized in GRBs because they describe emission episodes that most closely conform to the traditional definition of a pulse. FREDs are generally smooth, simple, monotonically-increasing and decreasing {\em monotonic pulses}. In fact, modeling approaches have almost exclusively assumed that all smooth pulses are monotonic ({\em e.g.,} \cite{1996ApJ...459..393N, 2005ApJ...627..324N, 1996ApJ...469L.109S, 2000ApJS..131....1L, 2000ApJS..131...21L, 2003ApJ...596..389K, 2012ApJ...744..141B, 2012MNRAS.419.1650N}), with shapes ranging from symmetric (equal rise and decay times) to highly asymmetric (short rise and long decay times). GRB pulses having longer rise than decay times are so rare that the \cite{2005ApJ...627..324N} model does not account for them. 

Monotonic models have been successfully applied to FRED-like pulses belonging to GRBs of different duration classes \citep{1986ApJ...301..213N, 1996ApJ...459..393N, 1997ApJ...489..175P, 2005ApJ...627..324N, 2011ApJ...735...23N, 2014ApJ...783...88H, 2018ApJ...855..101H, 2018ApJ...863...77H}, x-ray flares seen in the GRB afterglow phase \citep{2010MNRAS.406.2149M, 2016LPICo1962.4069H}, and bursts detected by instruments ranging from BATSE to Suzaku to Swift to Fermi GBM. The approaches have been more successful describing faint smooth pulses than bright complex ones.

Most GRBs exhibit a bright rapidly-varying component ({\em e.g.,} \cite{2001ApJ...552...57R}) that is separate from background noise. It can be difficult to separate this variable component from noise in the low photon count environments in which gamma-ray burst detectors operate. However, in bright GRBs variability can significantly exceed noise levels. 

Over the years, many authors have attempted to describe how variability (or {\em structure} following \cite{2018ApJ...863...77H}) might develop in GRB lightcurves. These have included models in which prompt emission is produced within internal shocks ({\em e.g.,} \cite{1994ApJ...427..708P, 1994ApJ...430L..93R, 1997ApJ...490...92K, 1998MNRAS.296..275D, 2009A&A...498..677B}), via relativistic turbulence ({\em e.g.,}\cite{2009MNRAS.395..472K, 2009ApJ...695L..10L, 2009MNRAS.394L.117N}), in mini-jets ({\em e.g.,}\cite{2003astro.ph.12347L, 2004ApJ...607L.103Y, 2014ApJ...782...92Z}), in fluctuations in black hole hyperaccretion disks ({\em e.g.,}\cite{2016MNRAS.463..245L}), and from fluctuations in the magnetic fields of accretion disks \citep{2016MNRAS.461.1045L}. However, the lack of a consistently-applied variability definition in these studies, along with the lack of a consistent pulse definition, means that many properties attributed to variability in one study were attributed to pulses in another study, and {\em vice versa}.

Pulse modeling is made difficult not only by instrumental noise and by the different pulse fitting algorithms used, but also by attempts to apply models to data that the algorithms do not adequately fit. The data used in various studies have been collected from instruments having different threshold sensitivities, spectral ranges and resolutions, and temporal resolutions. Flux that is not recognized as belonging to a fitted pulse has been treated variously as structure, as belonging to other (unfitted) pulses, as instrumental noise, or as some combination of these. The treatment of pulses has even been inconsistent within individual studies, especially when data from multiple instruments have been examined. In order to overcome these issues, a meaningful delineation between a monotonic pulse and its structure must be found, and this requires identification of some universal relationship that can be used to separate the two components.

\subsection{Time-reversed Structure} \label{sec:time-rev-intro}

In FRED pulses, where the underlying monotonic pulse is easy to model and subtract out, structure is found to take the form of a three-peaked wave structure in the residuals \citep{2014ApJ...783...88H}. The three peaks of the wave occur during the pulse rise, the pulse maximum, and the pulse decay, such that the distribution of peaks aligns with the asymmetric shape of the underlying pulse. The features are clearly seen in the residuals once the underlying smooth pulse has been fitted and removed using a model such as that of \cite{2005ApJ...627..324N}. These peaks are also associated with spectral evolution of the pulse. The monotonic components of a FRED pulse undergoes hard-to-soft evolution while the overall pulse (monotonic pulse plus structure) experiences a re-hardening at the time of each peak \citep{2015ApJ...815..134H}.

Structure is not only a component of simple FRED pulses; it is also found in GRB pulses associated with more complex emission episodes. Because pulses identified in high signal-to-noise ratio data typically contain more structure than those identified in low signal-to-noise data, brighter, more complex GRB pulses provide better insights into the nature of structure than do fainter, simpler FRED pulses \citep{2018ApJ...863...77H}. Structure is washed out at low signal-to-noise ratios, even when the underlying smooth pulse is not.

A study of bright GRB pulse light curves shows that most exhibit {\em time-reversible} characteristics \citep{2018ApJ...863...77H}. These strange characteristics occur for both the smooth and structured pulse components.

Monotonic pulse components are all to first order time-reversible. Pulse decay is the reverse process of pulse rise, with pulse peak intensity providing the delineation between the two processes. The main thing preventing a pulse's decay from appearing to be a time-reversed version of the pulse rise is its asymmetry. However, an adjustment to the rise time made by stretching it can cause the rise and decay to appear similar. The similarity can be seen by folding the stretched pulse light curve at the time of the pulse peak and aligning the folded rise profile with the decay profile.

\begin{figure}[ht!]
\begin{center}
\plottwo{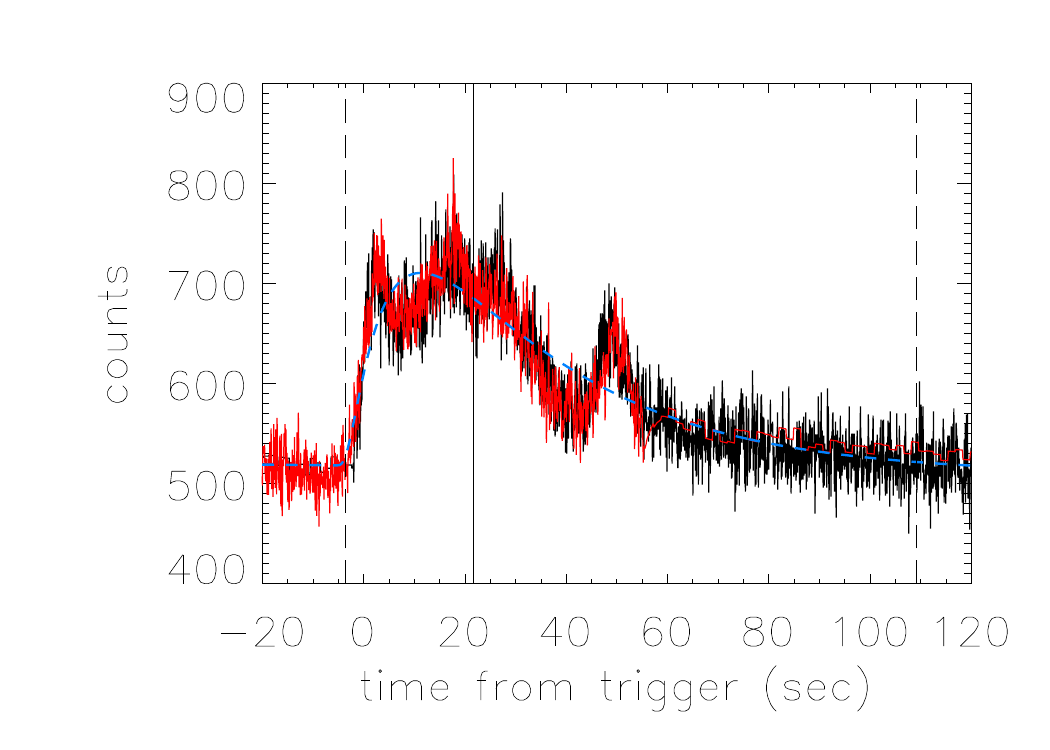}{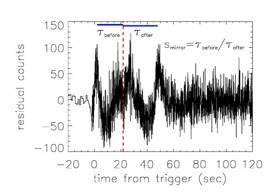}
\caption{Example of fitting a GRB pulse and its time-reversed structure. The left-hand panel demonstrates a smooth, asymmetric fit (dashed blue line) to the BATSE 64-ms 4-channel counts data of GRB pulse 659 (solid black line) using the \cite{2005ApJ...627..324N} model. The right-hand panel demonstrates the time-reversed fit to the residual pulse structure (solid black line) that is obtained by reversing in time the structure prior to the time of reflection (dashed red line) and matching it to the structure following the time of reflection after stretching it by an amount $s_{\rm mirror}$, such that time intervals align. This pulse, BATSE 659, is an example of the rollercoaster pulse morphology.
\label{fig:time-rev structure}}
\end{center}
\end{figure}

GRB pulse structure shares similar time-reversed properties to the monotonic pulse. Structure occurring prior to a {\em time of reflection} ($t_{\rm mirror}$) appears in reverse order following it \cite{2018ApJ...863...77H,2019ApJ...883...70H,2021ApJ...919...37H} and Liu and Zou (2023, preprint). This reverse ordering of features may also be accompanied by a temporal expansion (stretching) such that identifiable features found in the structure prior to $t_{\rm mirror}$ (denoted by the function $f(t_{\rm mirror}-t$)) match closely with those following it after accounting for a temporal stretching $s_{\rm mirror}$. In other words, $f(t_{\rm mirror}-t) \approx s_{\rm mirror} f(t-t_{\rm mirror}))$. This effect is demonstrated in the right panel of Figure \ref{fig:time-rev structure}. The time of reflection generally occurs near or just after the time that the monotonic light curve peaks; this time difference is referred to as the {\em offset}. \cite{2021ApJ...919...37H} finds that $86\%$ of BATSE GRB pulse light curves can be fitted by a model of a monotonic smooth pulse combined with a time-reversed structured component.

Time-reversed structure is demonstrated in Figure \ref{fig:time-rev structure}. Here, the data shown in the left panel (solid black line) have been fitted by an asymmetric monotonic pulse model (\cite{2005ApJ...627..324N}; shown as a dashed blue line), and the residuals of this fit shown in the right panel have been fitted with a time-reversed and stretched version of the same data to produce a combined model (solid red line) in the left panel. The process is described in detail in \cite{2021ApJ...919...37H}.

The amount of stretching in the time-reversed structure correlates with the amount of monotonic pulse stretching associated with the pulse asymmetry. Using the \cite{2005ApJ...627..324N} pulse model in which symmetric pulses have $\kappa=0$ and asymmetric pulses have $\kappa=1$, the stretching of the temporal structure is found to closely align with the pulse asymmetry.  \citep{2018ApJ...863...77H} find that when the pulse profile is symmetric, the time-reversed structure is distributed symmetrically around the time of reflection. In other words, $s_{\rm mirror} \approx 1-\kappa$. The $p-$value of this anticorrelation is $4.7 \times 10^{-20}$ for a sample of 160 BATSE GRB pulses for which $s_{\rm mirror}$ can be measured. This relationship demonstrates that the structure and monotonic pulse characteristics are related, and by linking them together it is clear that both components are associated with the same event. This relationship can be used to {\em define} GRB pulses as events having monotonic and structured components, both of which are time-reversed and stretched \citep{2018ApJ...863...77H, 2019ApJ...883...70H, 2021ApJ...919...37H}.

\begin{figure}[ht!]
\begin{center}
\includegraphics[width=0.5\textwidth]{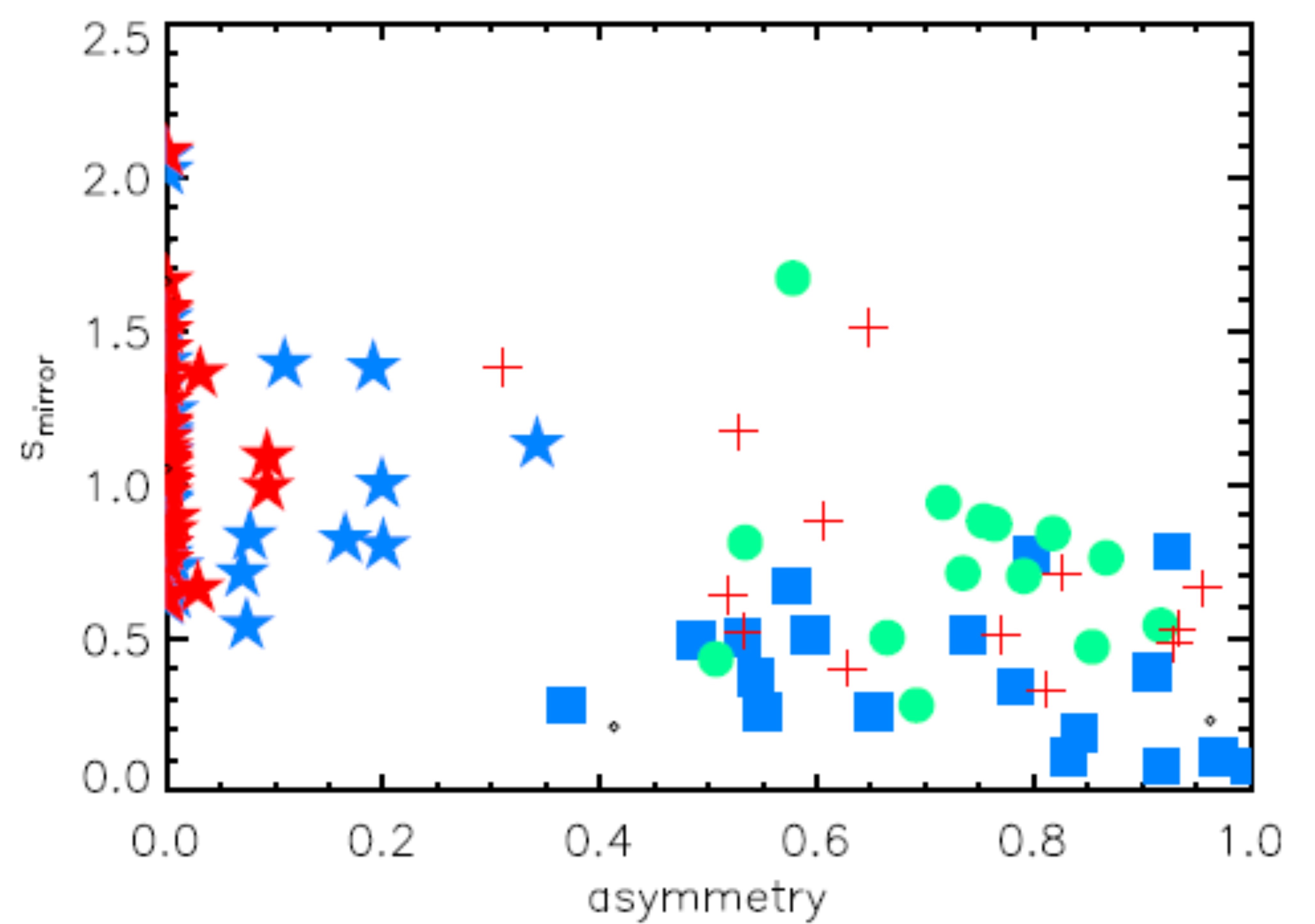}
    \caption{Morphological GRB pulse classes extracted from an unbiased set of bright BATSE GRBs \citep{2021ApJ...919...37H} Pulses exhibit similar light curves based on relative values of asymmetry $\kappa$ and $s_{\rm mirror}$. FRED pulses (blue squares; example is shown in the left panel of Figure \ref{fig:asymmetric_pulses}) have large $\kappa$ values and small $s_{\rm mirror}$ values. Rollercoaster pulses (green circles; example is shown in the left panel of Figure \ref{fig:time-rev structure}. Asymmetric u-pulses (red crosses; example is shown in the right panel of Figure \ref{fig:asymmetric_pulses}) have large $\kappa$ values and intermediate $s_{\rm mirror}$ values. U-pulses (red stars; example is shown in the right panel of Figure \ref{fig:symmetric_pulses}) have near zero $\kappa$ values and intermediate- to large-$s_{\rm mirror}$ values. Crown pulses (blue stars; example is shown in the left panel of Figure \ref{fig:symmetric_pulses}) have near zero $\kappa$ values and intermediate- to large-$s_{\rm mirror}$ values.
\label{fig:pulse classes}}
\end{center}
\end{figure}

Despite the pronounced anti-correlation between $s_{\rm mirror}$ and $\kappa$, individual pulses are widely dispersed from the $s_{\rm mirror} = 1-\kappa$ relationship. This is demonstrated in Figure \ref{fig:pulse classes} (data from \cite{2021ApJ...919...37H}).  This dispersion is not the result of uncertainty or noise in the measurements, but is instead produced by morphological variations in GRB light curves \cite{2021ApJ...919...37H}. 

\begin{figure}[ht!]
\begin{center}
\plottwo{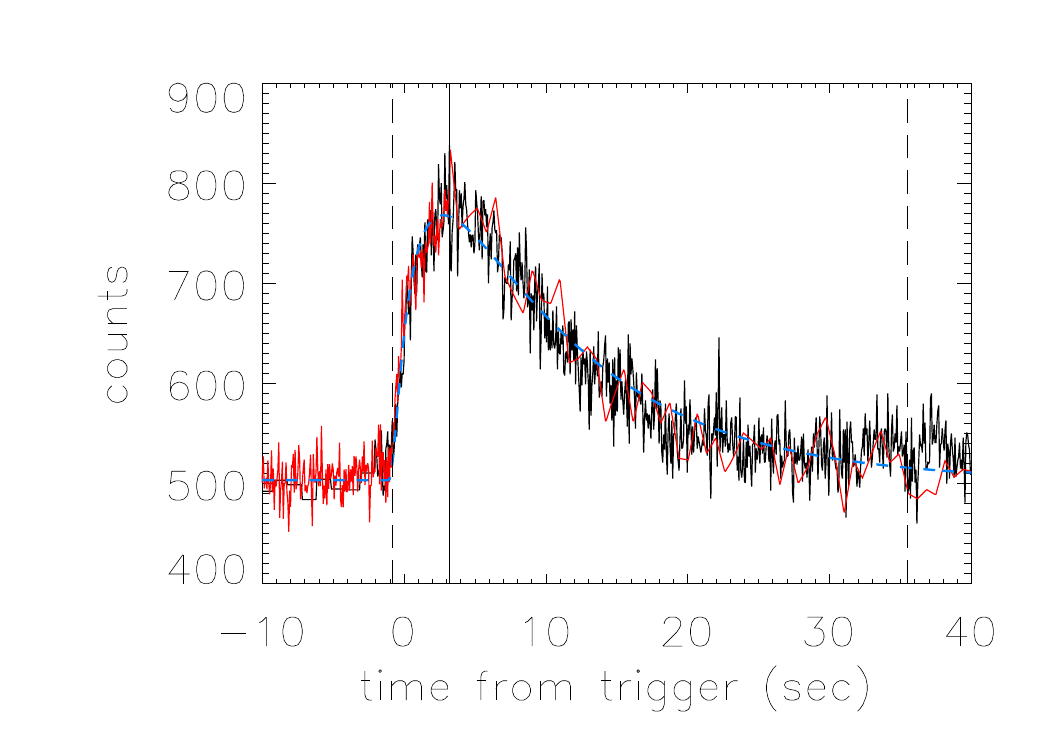}{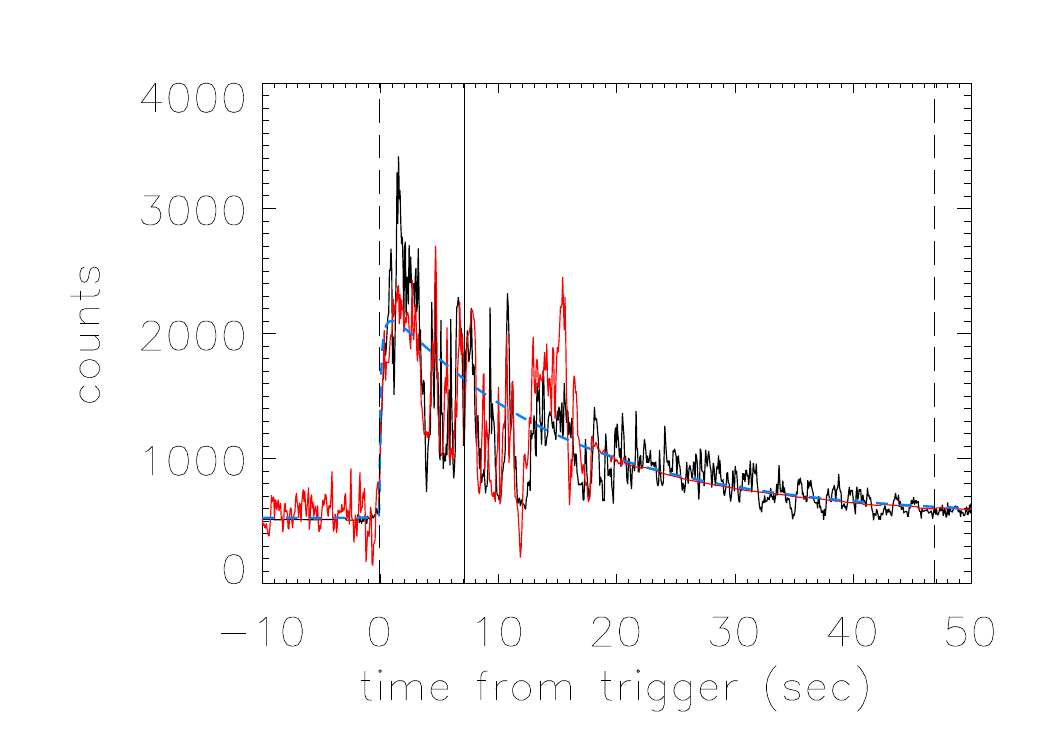}
\caption{Examples of asymmetric GRB pulse morphologies. BATSE 332 (left panel) is an example of a FRED pulse. BATSE 678 (right panel) is an example of an asymmetric u-pulse. BATSE 659 (Figure \ref{fig:time-rev structure}) is an example of a rollercoaster pulse.
\label{fig:asymmetric_pulses}}
\end{center}
\end{figure}

\begin{figure}[ht!]
\begin{center}
\plottwo{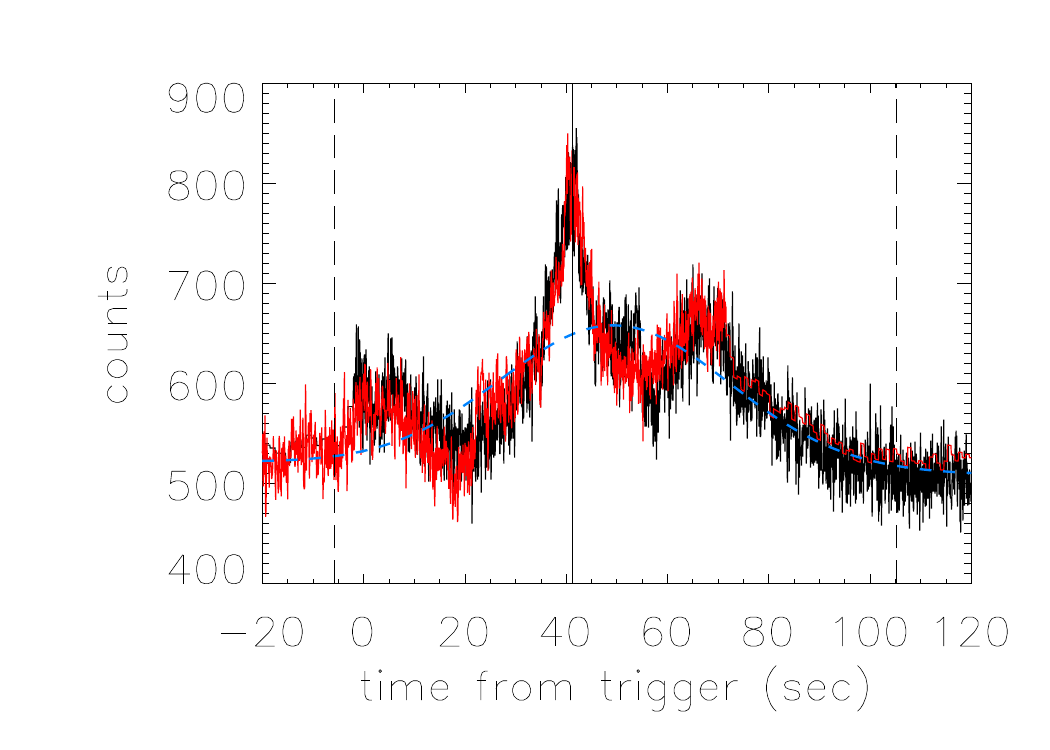}{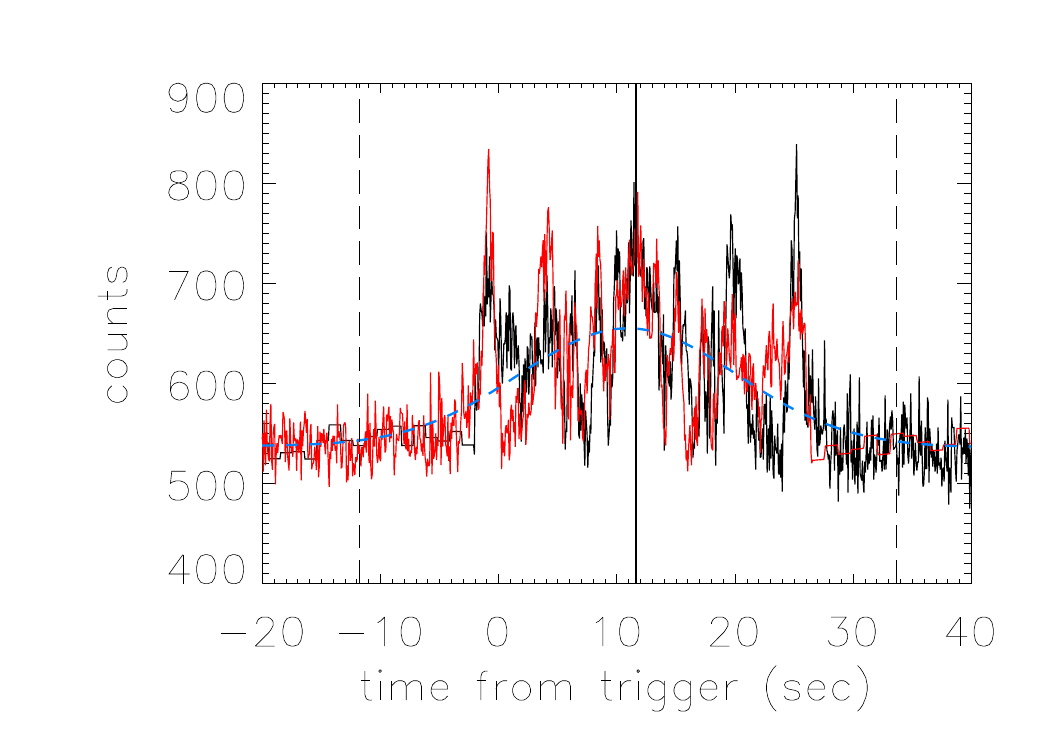}
\caption{Examples of symmetric GRB pulse morphologies. BATSE 351 (left panel) is an example of a crown pulse. BATSE 594 (right panel) is an example of an u-pulse.
\label{fig:symmetric_pulses}}
\end{center}
\end{figure}

The aforementioned morphological types, along with their associated time-reversed characteristics, appear to be as common among pulses from the short GRB class as they are among long GRB pulses. This has been somewhat difficult to ascertain due to the smaller number of photons received from short GRBs. However, FREDs can be clearly identified in short GRBs, and u-pulses were first recognized as a morphological pulse type in a study of short GRBs \citep{2018ApJ...855..101H}. Despite the smaller number of short GRB pulses that can be fitted by the time-reversed model, similar mechanisms are likely responsible for producing pulses in both GRB classes.


GRB emission is generally assumed to occur within relativistic jets launched from newly-formed black holes. Observations supporting jet models have included breaks in x-ray and optical afterglow lightcurves, as these indicate that GRBs contain highly beamed emission from relativistic jets \citep{1999ApJ...525..737R,1999ApJ...519L..17S}. Distances obtained from host galaxy redshift indicate that GRBs release $10^{48}$ to $10^{52}$ ergs \citep{2001ApJ...562L..55F, 2001ApJ...560L..49P, 2003ApJ...590..379B, 2008ApJ...685..354L, 2009ApJ...698...43R, 2010ApJ...711..641C} of energies on timescales of milliseconds to hundreds of seconds. Taken together, these observations have led to a standard model in which GRB prompt emission occurs within relativistic jets emanating from their progenitors (merging neutron stars for the short GRB class and collapsars for the long GRB class). GRB jets are generally assumed to point towards the observer.


Few attempts have been made to incorporate the observed time-reversed structure of GRB pulse light curves into standard jet models. Simply put, this is because standard radial jet models do not naturally produce these features. Although models involving forward and reverse shocks seem tailor-made for producing time-reversed structures, they have only been shown to do so in the simplest case of one forward and one reverse event \citep{2018ApJ...863...77H}. It is not obvious how they produce complex time-symmetric light curves. 
Additional types of models have been developed in the framework of bulk kinetic motion occurring within the confines of radially-evolving jets. The models involve impactors of optically-thick material that travel through the jet and interact with other blobs of material. Time-reversed pulse structure might be caused by a reversal of the impactor's motion, causing it to pass through blobs in reverse order \citep{2018ApJ...863...77H}, impactors moving through symmetrically-distributed blobs structures within the jet \citep{2018ApJ...863...77H}, symmetrically-structured impactors moving through simple blobs within a jet \citep{2018ApJ...863...77H}, impactors moving through density variations in a collapsing accretion disk \citep{2019IAUS..346..459H}, or density fluctuations transitioning between subluminal and superluminal motion in an expanding relativistic jet \citep{2019ApJ...883...70H}. These models all involve some {\em ad hoc} assumptions in order to be consistent with the observations.

The inconsistency in kinematic models to match the data appears to arise from the assumption that GRB jets point unwaveringly in the direction of the observer. This constraint places a difficult burden on models by requiring that all time-reversed mechanisms occur within a rapidly-expanding radial jet framework.

\section{Lateral Motion of Relativistic Jets and Gamma-ray Burst Pulse Light Curves} \label{sec:lateral} 

In order to examine time-reversed GRB pulse structure from a different perspective, we propose to relax the constraining assumption that GRB jets always point at the observer, and hypothesize that GRB jets move laterally while they expand rather than remaining fixed in place. Allowing the observer to see the jet cross-section as it sweeps across the line-of-sight forms the basis of the lateral jet motion hypothesis.
We hypothesize that this lateral motion is due to motion of the jet nozzle relative to the observer. This motion could be due to jet precession, chaotic motions as the nozzle seeks a stable configuration, or relative motion of the nozzle's frame relative to the observer.

\subsection{Lateral Jet Motion and Time-reversed Pulse Properties} \label{sec:time-rev} 

\begin{figure}[ht!]
\begin{center}
\includegraphics[width=0.5\textwidth]{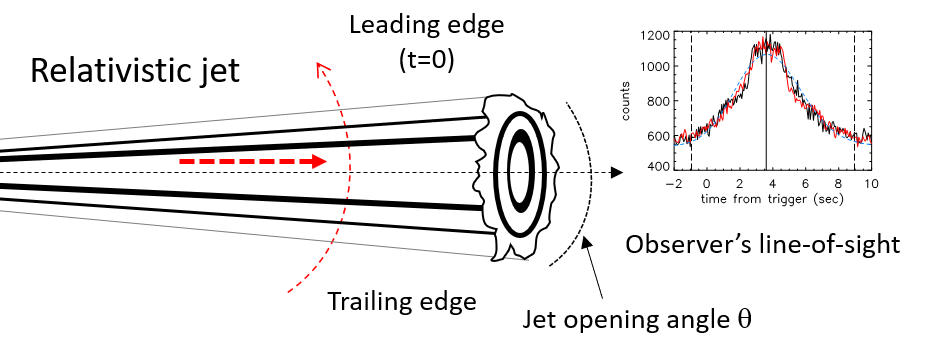}
    \caption{A simple model of a relativistic jet sweeping across an observer's line-of-sight. The jet expands outward with Lorentz factor $\Gamma$ (red dashed line), emitting light from the jet core and the surrounding structure as the jet sweeps (dotted arc) across the observer's line-of-sight (dotted line) with observed angular width $1/\Gamma$. Emission from the structured parts of the jet appear faint relative to the core due to relativistic beaming effects. An example of what the time-reversed light curve might look like is demonstrated using the light curve from a BATSE crown pulse light curve shown (from \cite{2021ApJ...919...37H}). 
\label{fig:jet structure}}
\end{center}
\end{figure}

A radially-symmetric jet sweeping laterally across an observer's line-of-sight can naturally account for time-reversed GRB pulse light curve structures. When a jet crosses the line-of-sight, structures surrounding the jet are both the first and the last things to be seen by the observer. If a pulse begins when the edge of the jet first crosses the observer's line-of-sight, then it ends as it exits. Any radial structures within the jet will also appear as time-reversed features in the light curve, as shown in Figure \ref{fig:jet structure}.

The jet structure thus likely represents regions producing more emission or less emission than that emitted by the central core of the jet. Without specifying the emission mechanism, we can naively assume that photon production is proportional to the amount of ejecta emitted in specific regions of the jet, or perhaps to the efficiency with which photons are produce in these regions. Thus, structure could represent jet regions in which ejecta production inordinately productive, such as in mini-jets ({\em e.g.,} \cite{2006MNRAS.369L...5L, 2009ApJ...695L..10L, 2009MNRAS.394L.117N, 2016MNRAS.459.3635B, 2016ApJ...816L..20G, 2024MNRAS.527.12178}). Conversely, they might represent regions where mechanisms producing jet ejecta are less productive, such as regions along the edge of the jet where it is less efficient at burning through its progenitor star, or where less material is flowing outward, such as in a jet's turbulent cocoon. Regardless of the mechanism that produces pulse structure, this model assumes that the jet contains general asymmetries that are axisymmetric.

The bulk motion in a relativistic jet is characterized by Lorentz factor $\Gamma$, where
\begin{equation}
\Gamma=\frac{1}{\sqrt{1-\beta^2}},
\end{equation} 
and where $\beta=v/c$ and $c$ is the speed of light. GRBs have highly relativistic jets, with typical $\Gamma\approx 300$.
As a jet crosses the line-of-sight, the jet's cross-section is observed, with several relativistic effects coming into play. First, the observer sees the jet emission confined to a narrow cone of angular diameter $1/\Gamma$.
Second, the intensity in the beam is Doppler boosted by an amount $D^2$, where the Doppler factor $D$ is
\begin{equation}
D=\frac{1}{\Gamma(1-\beta \cos \theta)}.
\end{equation}
The {\em headlight effect} causes emission from the line-of-sight to outshine that from the rest of the jet. Thus, as the jet laterally traverses the line-of-sight, each contour of the jet is captured at a specific time as part of a tracing on the light curve.
Third, the pulse duration $\tau$ indicates the time it takes for the jet to be seen crossing the line-of-sight. Both the opening angle of the jet and the rate at which the jet moves across the line-of-sight contribute to the duration.
Fourth, transverse motion across the line-of-sight provides an alternate explanation for pulse asymmetry and hard-to-soft evolution that is not limited to radial events within the jet. Lateral jet motion might help explain differences between symmetric pulses (crown pulses and u-pulses) and asymmetric pulses (FREDs, rollercoaster pulses, and asymmetric u-pulses).

\begin{figure}[ht!]
\begin{center}
\includegraphics[width=0.5\textwidth]{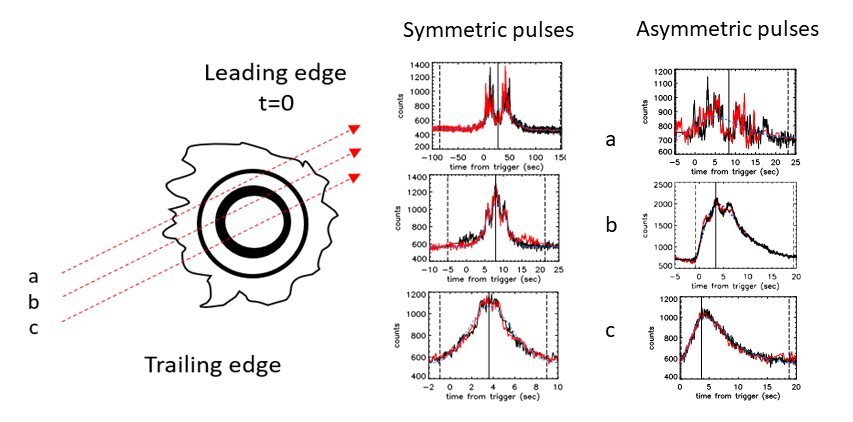}
\caption{Pulse light curve morphology as a function of off-axis viewing. An observer can see a slightly
different cross-section of the jet structure depending on where the jet crosses the line-of-sight (left column), as relativistic beaming causes different features in symmetric pulses (center column) or asymmetric pulses (right column) to be magnified along each path (a, b, c). Depending on pulse asymmetry, (a) when the edge but not the core of the jet cross the line-of-sight, then an observer may see a u-pulse (a-center) or an asymmetric u-pulse (a-right), (b) when the jet is viewed slightly off-axis, then an observer may see a crown/u-pulse (b-center) or a rollercoaster pulse (b-right), or (c) when the jet core crosses the line-of-sight such that the core brightness overpowers the edge, then the observer might see a crown pulse (c-center) or a FRED (c-right). The sample light curves shown are from symmetric pulses observed by BATSE and described by \cite{2021ApJ...919...37H}. In these light curves, data (black lines) are overlaid by the time-reversed model (red lines), which have been folded and stretched around the time of reflection (central dashed vertical line) between the boundaries of the pulse duration (other vertical dashed lines).
\label{fig:lightcurves}}
\end{center}
\end{figure}

 Slightly different features will be seen when the edge of a jet crosses the line-of-sight than when the jet core crosses it. An off-axis pass causes an observer to see a slightly different slice of the jet's emission profile. This process is demonstrated in Figure \ref{fig:lightcurves}, where the observer sees a slightly different profile depending on where the jet direction intercepts the line-of-sight. Since most pulses can be characterized by only a few morphological types \cite{2021ApJ...919...37H}, the jet structure must be fairly similar from GRB to GRB. 
 
Figure \ref{fig:pulse classes} demonstrates that structured GRB pulses fall into a remarkably small number of distinctly-different light curve morphologies. If these morphological variations are the result of jet profiles viewed from different off-axis locations at different evolutionary times, then GRB jets must have very similar makeups that potentially do not change much as jets expand. Armed with this assumption, we can study GRB jet structure by considering each pulse to be a cross-sectional structure of a jet.

The idea that a pulse light curve might be used to trace jet structure is not new. This approach has been used to map out pulsar jets from light curves using orbital precession in the binary pulsar system PSR $1913+16$ \citep{1989ApJ...347.1030W}. Although GRBs do not repeat, there is great potential of using their light curves to trace out structure that seems to bear remarkable morphological similarities from jet to jet \citep{2021ApJ...919...37H}.

\subsection{Lateral Jet Motion and Pulse Asymmetry} \label{asymm} 

Jet models must address more characteristics than just light curve similarities. They need to also address the variation in pulse asymmetries, as well as the hard-to-soft spectral evolution generally observed in GRB pulses.

Half of all observed GRB pulses have time-asymmetric profiles with faster rise than decay times (e.g., \cite{1994ApJ...423..432N,2021ApJ...919...37H}). FRED pulses, rollercoaster pulses, and asymmetric u-pulses are examples of asymmetric pulse types. Besides the asymmetry of the underlying smooth pulse, the distribution of structure around the time of reflection is also asymmetric. Asymmetry in the light curve's structure is characterized not only by an increased temporal separation between light curve features but also by a broadening of these features \citep{2018ApJ...863...77H, 2019ApJ...883...70H, 2021ApJ...919...37H}. Thus, in time-asymmetric pulses, a relatively abrupt transition delineates the temporal compression of events preceding the time of reflection from those broader time-reversed events following it. 

One way by which an observer may see an asymmetric jet profile is if the jet is bent rather than straight as it crosses the line-of-sight. In order to allow for this, viable models must consider how jet motion across the line-of-sight might cause it to bend. This has to do with jet structure and collimation.

There are three main types of helical jets ({\em e.g.,} \cite{1997VA.....41...71S}): {\em ballistic helical jets} (where the flow of ejecta is along a straight line but where the ejecta direction changes), {\em helically bent jets} (where the flow of ejecta is along a curved jet axis), and {\em internal helical jets} (where helical structure exists within an otherwise straight jet). For our modeling we consider the first {\bf two of these jet types}.

\section{Model 1. Helically-bent rigid jet model} \label{sec:rigid}

\begin{figure}[ht!]
\begin{center}
\includegraphics[width=0.5\textwidth]{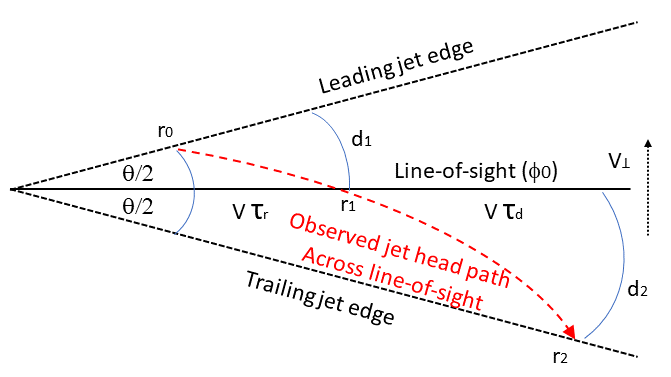}
\caption{A rigid/semi-rigid jet model capable of explaining time-reversed GRB pulse structure. The model also shows how asymmetry can be introduced into the pulses created from the rapid expansion of a laterally-moving rigid/semi-rigid jet. As the jet moves across the line-of-sight from the leading edge to the jet center, it expands from distance $r_0$ to $r_1$ during the pulse rise duration $\tau_r$, where $r_1 = r_0 + v \tau_r$ (corresponding to having the jet traverse a lateral distance $d_1$ at transverse velocity $v_\perp$). As the jet moves from the jet center to the trailing edge during the pulse decay duration $\tau_d$, it expands from distance $r_1$ to $r_2$, where $r_2 = r_0 + v \tau_r + v \tau_d r_0 + v \tau$ (simultaneously traversing a lateral distance $d_2$ at velocity $v_\perp$).
\label{fig:jet_annotated}}
\end{center}
\end{figure}

Consider a rigid or semi-rigid jet that is able to retain its structure while moving laterally across the line-of-sight. Such a jet is a form of the helically-bent jet model described above. Geometric jet constraints are present that do not depend on either the mechanism that contains the jet or the mechanism responsible for causing the jet’s lateral motion. We demonstrate here that rigidity constrains the jet to cross the line-of-sight close to the nozzle where the jet ejecta has been emitted by the progenitor. Figure \ref{fig:jet_annotated} shows how a rigid/semi-rigid jet with opening angle $\theta$ expands as it crosses the line-of sight starting at distance $r_0$ from the source.

We define the pulse duration $\tau$ as the time it takes for the observable jet to cross the line-of-sight. For this model, we assume that the jet must cross at a distance equal to its diameter $d$ as it moves at lateral velocity $v_\perp$ in order that an observer see the full jet cross-section. Defining the {\it crossing distance} $r_{\rm max}$ as the maximum distance from the source at which the trailing edge of the jet crosses the line-of-sight, we have $r = v_\perp \tau / \theta$, where $\theta$ is the jet opening angle. Because the physical upper limit is $v_\perp < c$,

\begin{equation} \label{eq:cross}
     r_{\rm max} < c \tau / \theta,
\end{equation}
or
\begin{equation}     
     r_{\rm max}{\rm(cm)} < 1.7\ \times 10^{12} \tau / \theta(^\circ).
\end{equation}

\begin{figure}[ht!]
\begin{center}
\includegraphics[width=0.5\textwidth]{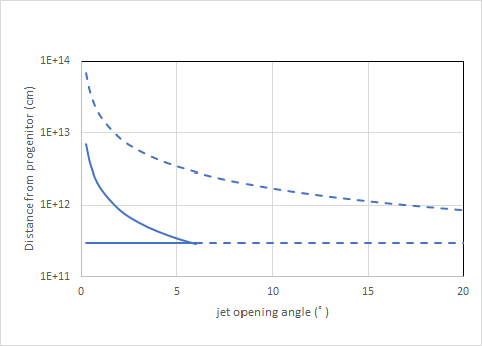}
\caption{Simple theoretical distance limits within which a rigid/semi-rigid relativistic jet can cross the line-of-sight to produce a 10 s GRB pulse having time-reversed properties. The curved dashed line represents the maximum distance from the nozzle at which the jet can cross the line-of-sight and correspond to a lateral velocity of $v_\perp = c$. The curved solid line represents a maximum crossing distance corresponding to $v_\perp = 0.1 c.$ The straight solid/dashed line corresponds to the minimum radial length the jet can have and still cross the line-of-sight during its near-lightspeed expansion.
\label{fig:crossing}}
\end{center}
\end{figure}

For a GRB pulse with duration $\tau = 10$ s and for jets having opening angles in the expected range $1^\circ \le \theta \le 10^\circ$, crossing distances are in the range $10^{12} \le r_{\rm max} \le 10^{14}$ cm (dashed curved line in Figure \ref{fig:crossing}. These crossing distances are too small by at least an order of magnitude to be reconciled with the standard GRB emission mechanisms that predict internal shocks to be produced at distances of $r = 10^{14}$ to $10^{15}$ cm from the source (citation). These limits are further constrained because the total velocity $v_{\rm tot}$ of the jet's {\em emitting region} from which photons are produced, must be less than c. In other words,
\begin{equation}
    v_{\rm tot} = \sqrt{v^2 + v_\perp^2 - \frac{v^2 v_\perp^2}{c^2}} \le c.
\end{equation}

For more reasonable limits on the lateral jet velocities ($v_\perp << c$), crossing distances must be be even closer to the progenitors (solid curved line in \ref{fig:crossing}). For a $\tau = 10$ s pulse, $v_\perp = 0.1c$ reduces the crossing distance to $10^{11} \le r_{\rm max} \le 10^{13}$ cm. 

The minimum crossing distance is defined by the radial expansion of the rigid/semi-rigid jet. The minimum jet length at which the jet can cross the line-of-sight to produce a time-reversed light curve is characterized by the expansion velocity of the jet. If the jet starts crossing the line-of-sight the moment it is launched, then it will have expanded to a crossing distance $r_{\rm min} = v_\perp \tau$ by the time it completes its crossing. This minimum crossing distance is defined by the lower solid/dashed straight line in Figure \ref{fig:crossing}.

\begin{figure}[ht!]
\begin{center}
\includegraphics[width=0.5\textwidth]{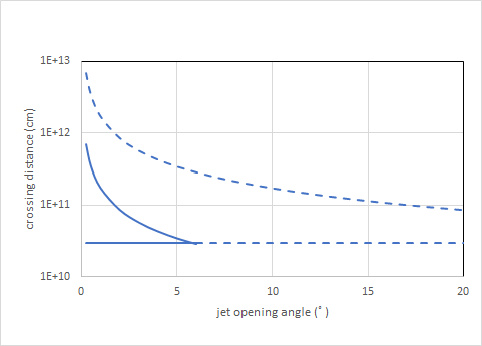}
\caption{Simple theoretical distance limits within which a relativistic jet can cross the line-of-sight to produce a 1 s GRB pulse having time-reversed properties. 
\label{fig:crossing1}}
\end{center}
\end{figure}

\begin{figure}[ht!]
\begin{center}
\includegraphics[width=0.5\textwidth]{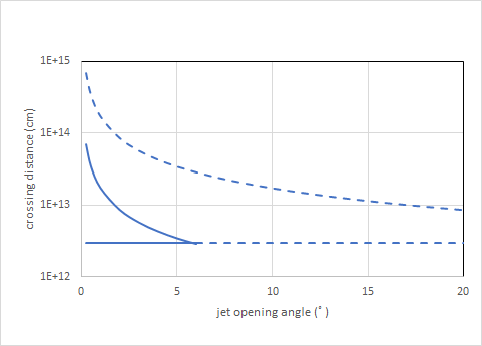}
\caption{Simple theoretical distance limits within which a relativistic jet can cross the line-of-sight to produce a 100 s GRB pulse having time-reversed properties.
\label{fig:crossing100}}
\end{center}
\end{figure}

Figure \ref{fig:crossing1} shows the same crossing distance parameter space as is shown in Figure \ref{fig:crossing}, except for GRB pulses of $\tau = 1$ s. Simiarly, Figure \ref{fig:crossing100} shows the same crossing distance parameter space as is shown in Figure \ref{fig:crossing}, except for GRB pulses of $\tau = 100$ s.

These figures show that it is geometrically possible for a laterally-moving rigid jet to produce pulses of light having time-reversed structures consistent with those of GRB pulse light curves. However, the lateral velocities of these jets must be relatively large in order to cross the observer's line-of-sight on the relatively short timescales of GRB pulses. Since relativistic lateral jet lateral velocities seem unlikely, the jet opening angles must be similarly constrained to $\theta \le 5^\circ$. For a typical GRB pulse, these constraints also require the jet to cross the line-of-sight within a relatively narrow range of distances from the nozzle.

The ability of the rigid/semi-rigid jet model to explain pulse asymmetry depends on the rigidity of collimated jet and thus on how the jet moves laterally. For example, a rigid jet with a lateral velocity of the form $v_\perp = v_{\perp 0} r$ (where $v_{\perp 0}$ is a constant) will move across the line-of-sight at a constant angular velocity. The leading edge of the laterally-moving jet will take the same amount of time to cross the observer's line-of-sight as the trailing edge, producing a symmetric pulse rather than an asymmetric one.

If the jet is semi-rigid such that the lateral jet velocity is constant regardless of the distance from the central engine, then the jet head will lag behind the jet base as material flows within it. This semi-rigidity allows pulse asymmetry to form simply due to curvature accompanying jet expansion.

Figure \ref{fig:jet_annotated} demonstrates how pulse asymmetry may arise from a rigid or semi-rigid GRB jet crossing the line-of-sight. The expansion can cause the lateral range of the jet to be larger far from the central engine than near to it. If the jet is moving laterally with a constant velocity $v_\perp$, then the time $\tau_d$ needed for the trailing edge to cross relative to the jet center will be longer than the time $\tau_r$ for the jet center to cross relative to the leading edge.

The lateral distance $d_1$ crossed during the pulse rise is related to the pulse rise duration by
\begin{equation} \label{eq:d1a}
    d_1 = v_\perp \tau_r.
\end{equation}
Similarly, the lateral distance $d_2$ crossed during the pulse decay is related to the pulse decay duration by
\begin{equation} \label{eq:d2a}
    d_2 = v_\perp \tau_d.
\end{equation}
During the duration of the pulse rise, the jet expands at velocity $v$, and $d_1$ is also given by
\begin{equation} \label{eq:d1b}
    d_1 = (r_0 + v \tau_r) \theta/2.
\end{equation}
Also,
\begin{equation} \label{eq:d2b}
    d_2 = (r_0 + v \tau_r + v \tau_d) \theta/2.
\end{equation}
Note that the pulse duration $\tau$ is given by $\tau = \tau_r + \tau_d.$
Combining Equation \ref{eq:d1a} and Equation \ref{eq:d1b} yields 
\begin{equation} \label{eq:tau_r}
    \tau_r = \frac{r_0 \theta / 2}{v_\perp - v \theta / 2}.
\end{equation}
Similarly, combining Equation \ref{eq:d2a} and Equation \ref{eq:d2b} yields 
\begin{equation} \label{eq:tau_d1}
    \tau_d = \frac{(r_0 + v \tau_r) \theta / 2}{v_\perp - v \theta / 2}.
\end{equation}
Substituting Equation \ref{eq:tau_r} into Equation \ref{eq:tau_d1} results in 
\begin{equation} \label{eq:tau_d}
    \tau_d = \frac{v_\perp r_0 \theta/2}{(v_\perp - v \theta / 2)^2}.
\end{equation}
The pulse duration can be found in terms of $r_0$, $v$, and $v_\perp$ by combining Equation \ref{eq:tau_r} and Equation \ref{eq:tau_d}:
\begin{equation} \label{eq:tau}
    \tau = \frac{r_0 \theta/2 (2 v_\perp -v \theta/2)}{(v_\perp - v \theta / 2)^2}.
\end{equation}

The stretching factor $s$ is then approximately given as
\begin{equation}\label{eq:s}
    s \approx \tau_r/\tau_d = 1 - \frac{v \theta /2}{v_\perp}.
\end{equation} 

Equation \ref{eq:s} demonstrates that pulses are symmetric ($s=1$) when a jet expands only slightly relative to its motion across the line-of-sight, {\em e.g.,} when $v \ll 2 v_\perp / \theta$. Pulses are asymmetric ($s=0$) when a jet expands significantly relative to its motion across the line-of-sight, {\em e.g.,} when $v \approx 2 v_\perp / \theta$. For example, a jet expanding at $v \approx c$ and crossing with velocity $v_\perp \approx c/10$ will appear to be much more symmetric when the opening angle is small {\bf ($s \approx 0.1$ for $\theta = 1^\circ$)} than when the opening angle is large {\bf ($s \approx 0.9$ for $\theta = 10^\circ$)}.


\section{Model 2. Ballistic jet model} \label{sec:firehose}

\begin{figure}[ht!]
\begin{center}
\includegraphics[width=0.8\textwidth]{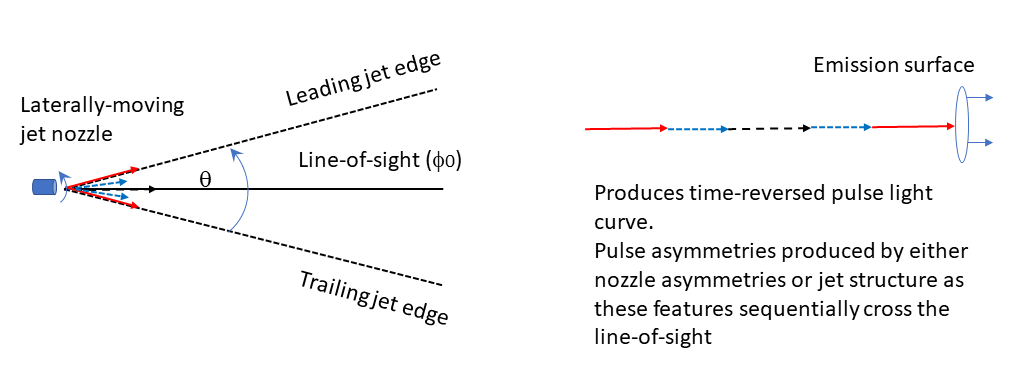}
\caption{A ballistic model in which a is jet launched from a moving axisymmetric nozzle (blue cylinder). As the jet crosses the line-of-sight, structure from different locations in the nozzle, represented by differing relativistic particle densities, compositions, and/or energies is sent in the observer's direction. Once these particles reach a certain distance, they interact to produce photons at an emission surface (denoted by the the blue ellipsoid. Because these events produce photons in the order they were emitted from the nozzle, the observer sees sequential emission from the leading edge of the laterally-moving jet (red solid line), the center of the jet (blue dotted line, black dashed line, blue dotted line), and the trailing edge of the jet (red solid line). 
\label{fig:moving_nozzle}}
\end{center}
\end{figure}

A ballistic helical jet is one in which the flow of ejecta is along a straight line but where the nozzle ejecting it changes direction. The jet itself is not rigid and thus does not move laterally, but particle emission from across the jet's cross section is funneled sequentially towards the observer as the nozzle moves across the line-of-sight. Unlike the rigid jet model, where the emitting region is forced to be close to the nozzle, the emitting region in the ballistic jet model can be at much larger distances, consistent with standard GRB shock models. As the particle stream interacts to produce photons, these photons are produced in a sequential order corresponding to the structure in the jet cross section. The pulse duration is a direct measurement of the time it takes for the nozzle to cross the line-of-sight, and time-reversed structure in the pulse light curve indicates radially-symmetric structure of the moving jet nozzle. A simple drawing of a jet launched from a moving nozzle is shown in Figure \ref{fig:moving_nozzle}.

\begin{figure}[ht!]
\begin{center}
\includegraphics[width=0.5\textwidth]{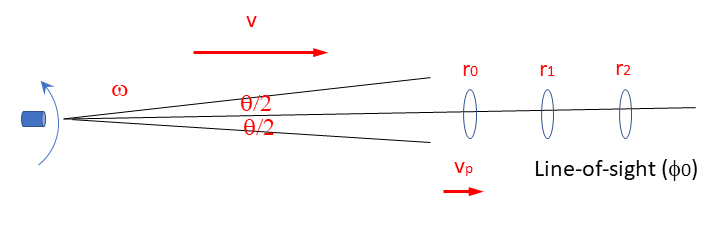}
    \caption{Asymmetry from ballistic jet expanding outward with velocity $v$. As the jet nozzle (blue cylinder) moves across the line-of-sight (angle $\phi_0$) with angular velocity $\omega$, the emitting region (oval) expands outward at velocity $v_p$. During the pulse rise phase, the nozzle rotates by angle $\theta/2$ and the emitting region expands from distance $r_0$ to distance $r_1$. During the subsequent pulse decay phase, the nozzle rotates again by $\theta/2$ and the emitting region expands from distance $r_1$ to distance $r_2$. Pulses appear asymmetric because it takes longer for ejecta to travel to the emitting region during the pulse decay phase than during the pulse rise phase.
\label{fig:firehose asym}}
\end{center}
\end{figure}

The ballistic jet model can explain some (but not all) of the pulse asymmetry naturally. This results from the natural process of having the jet emitting region expand with time as the jet pressure forces this region progressively farther from the nozzle. This effect is demonstrated in Figure \ref{fig:firehose asym}. We apply a simple estimate to determine the amount of pulse asymmetry that this introduces in the following fashion. 

After the jet ejecta has been launched at velocity $v$, it travels a distance $r_0$ before attaining the conditions necessary for a emitting region to form. Prior to this, the jetted material does not emit photons. As the emitting region emits, it is also pushed away from the nozzle at a velocity $v_p$. Assuming the jet has radial symmetry, the pulse peak corresponds to the time the jet center emits along the line-of-sight. During the pulse rise phase, the nozzle turns at angular velocity $\omega$ by an amount $\theta /2$, or half the opening angle, which also results in the emitting region expanding to distance $r_1$. During the pulse decay phase, the nozzle turns by an additional amount $\theta /2$ , which results in the emitting region expanding to distance $r_2$.

The time $t_0$ that it takes for the ejecta to reach $r_0$ is
\begin{equation}
    t_0 = v/r_0.
\end{equation}
The time $t_{01}$ that it takes for the emitting region to expand from $r_0$ to $r_1$ is
\begin{equation}
   t_{01} = \frac{\theta / 2}{\omega} = r_1 / v_p,
\end{equation}
such that 
\begin{equation}
    r_1 = \frac{\theta v_p}{2 \omega}.
\end{equation}
Similarly,
\begin{equation}
    r_2 = \frac{\theta v_p}{2 \omega}.
\end{equation}
Since the time it takes for the emission at the jet center to reach the emitting region $t_cs$ is 
\begin{equation}
    t_c = \frac{r_0 + r_1}{v} = t_0 + \frac{r_1}{v} = t_0 + \frac{\theta}{2 \omega} \frac{v_p}{v},
\end{equation}
the pulse rise time $\tau_{\rm rise}$ is given as
\begin{equation}
    \tau_{\rm rise} = t_{10} + t_c = \frac{\theta}{2 \omega} (1+\frac{v_p}{v}).
\end{equation}
The pulse decay time $\tau_{\rm decay}$ is similarly
\begin{equation}
    \tau_{\rm decay} = 2 t_{10} + t_c = \frac{\theta}{2 \omega} (1+\frac{2 v_p}{v}),
\end{equation}
such that the pulse duration $\tau$ is given by
\begin{equation}
    \tau = \tau_{\rm rise} + \tau_{\rm decay} = \frac{\theta}{\omega} (1 + 3 \frac{v_p}{v}).
\end{equation}
Similarly, the stretching factor $s$ is roughly
\begin{equation}
    s = \frac{\tau_{\rm rise}}{\tau_{\rm decay}} = \frac{1+v_p/v}{1+2 v_p/v}.
\end{equation}

Pulses are symmetric when the emitting region expands only slightly during the jet expansion, {\em e.g.,} when $v_p/v \rightarrow 0$, which leads to $s \rightarrow 1$. Similarly, pulses have the greatest asymmetry when the emitting region expands a large amount during the jet expansion, {\em e.g.,} when $v_p/v$ is large. If the emitting region does not accelerate, then the upper limit is $v_p/v \rightarrow 1$ which leads to $s \rightarrow 2/3$.
The asymmetry produced by expansion of the jet emitting region is therefore unable by itself to produce the small stretching factors of $0.1 \le s \le 0.6$ found in FREDs, rollercoaster pulses, and asymmetric u-pulses. However, it is interesting to note that a natural separation between symmetric and asymmetric pulse morphologies occurs at $\kappa \approx 1/3$ \citep{2021ApJ...919...37H}, which with our proxy of $\kappa \approx 1-s$ is consistent with symmetric pulse types having asymmetries that are due entirely to jet expansion.

Additional pulse asymmetry can be introduced if the jet changes according to some of the mechanisms proposed here:
\begin{itemize}
    \item The lateral motion of the nozzle can slow as it crosses the line-of-sight. This can cause the trailing part of the jet to take longer to cross than the leading part. Given that the jet also continues expanding during this time, this could produce a substantial amount of additional pulse asymmetry.
    \item The jet opening angle can increase with time. After the jet nozzle has ejected a large amount of material, it seems reasonable that the mechanisms causing the jet to be confined might eventually break down. These would result from the tapering off of material that feeds the accretion disk as well as the weakening of the toroidal magnetic field that collimates the jet. The loss of these constraints could weaken the jet collimation at the nozzle and cause $\theta$ to increase with time.
    \item Curvature of the jet head may cause photons emitted at off-axis locations to arrive at the observer later than those emitted on-axis ({\em e.g.,} \cite{2001ApJ...554L.163I}).
    \item The jet emitting region can accelerate outward as the jet crosses the line-of-sight instead of expanding at a constant velocity. This can increase the time it takes for jet ejecta to reach the emitting region during the pulse decay relative to the time taken during the pulse rise.
\end{itemize}   

We note here that not all asymmetric GRB pulses have easily-discernible structure at the end, which we would normally take as an indication that the trailing edge of the jet had crossed the line-of-sight. FRED and rollercoaster $s-$values are determined primarily from the triple-peaked structure that occurs around the middle of the pulse rise, at the pulse peak, and around the middle of the pulse decay \citep{2018ApJ...863...77H}. These pulses exhibit time-reversed structure occurring towards the center of the pulse rather than towards the end of the pulse. Thus, it may be that a slowly-decaying GRB pulse does not have well-defined structure at its end simply because the jet that does not finish crossing the line-of-sight until long after the main part of the jet has crossed. This is consistent with a slowing nozzle or an increasingly large opening angle (for a ballistic jet) or with a curved jet (for a rigid or semi-rigid jet).

As a jet lengthens, the surface area of the jet head increases and the volume into which jetted material flows becomes progressively larger. Thus, the energy density decreases as the jet expands. Furthermore, the average particle energy decreases as the jet pushes through and interacts with the surrounding external medium. Both of these effects will contribute to a graduate weakening of the spectrum produced at the emitting region, and this weakening should be observed in the overall hard-to-soft spectral evolution found in both GRBs and GRB pulses. We will explore spectral models produced by these models in a subsequent manuscript.

\section{Pulse Durations and Separations} \label{sec:separations} 

There is less confusion with identifying GRB pulses once time-reversed structure has been incorporated into the GRB pulse definition, since most variations found within GRBs do not qualify as individual pulses. In the case of bright bursts, most of emission can be accounted for by a small number of time-reversed pulses, reducing the possibility that faint, unfitted pulses may be present. Most of these bursts' emission is contained within only one or two pulses \citep{2018ApJ...855..101H, 2018ApJ...863...77H, 2021ApJ...919...37H}. 

X-ray flares, which are also scarce, have the characteristics of late GRB pulses that occur during the afterglow phase and after the prompt emission has ended.
X-ray flare intensities decrease, durations increase, and spectra soften with time ({\em e.g.,} \cite{2007ApJ...671.1903C, 2007ApJ...667.1024K}), similar to what is found for GRB pulses ({\em e.g.,} \cite{2000ApJ...539..712R, 2014ApJ...783...88H}. 

Separations between pulses and pulse durations can be used to place additional constraints on mechanisms capable of producing lateral jet motion. First, as mentioned above, pulses/flares are rare. Second, there is no indication that GRBs exhibit any form of periodic motion. Third, there is the aforementioned evolution of pulse properties with time.

If GRB pulses arise as a ballistic jet crosses the line-of-sight, then the angular velocity of the nozzle can be estimated from the pulse duration (excluding the asymmetry-producing jet expansion effects described in Section \ref{sec:firehose}) as $\omega \approx \theta/\tau_1$ where $\theta$ is the jet opening angle and $\tau_1$ is the duration of the initial pulse. Pulses produced via a periodic process would be expected to repeat after a time ${\rm sep} = 360^\circ/\omega$. Therefore, relative to the duration of the first pulse, the ratio {R} should be
\begin{equation}
    R={\rm sep}/\tau_1
\end{equation}
for all GRB that produce repeating pulses in this manner.
Similarly, the duration of the second pulse $\tau_2$ should be the same as $\tau_1$.

\cite{2021ApJ...919...37H} systematically studied pulses in a sample of BATSE GRBs using the time-reversed pulse definition which prevents pulses from being confused with pulse structure. Although most GRBs appear to be single-pulsed, this sample can also be used to study the evolution of pulse properties within multi-pulsed GRBs. Unfortunately, this robust sample cannot easily be redshift-corrected, as few redshift measurements were available for BATSE GRBs. However, we can use a normalization process to remove the interfering effects of redshift. By normalizing both the pulse separation $sep$ and the duration of the second pulse $\tau_2$ to the duration of the first pulse $\tau_1$, the redshift-independent properties $\tau_2/\tau_1$ and $sep/\tau1$ can be obtained. A comparison of these properties for the burst sample allows us to determine whether or not the duration of the second pulse changes as a function of pulse separation. 

\begin{figure}[ht!]
\begin{center}
\includegraphics[width=0.4\textwidth]{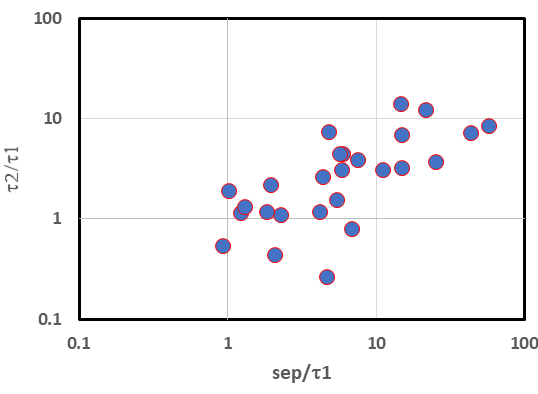}
\caption{Properties of multi-pulsed GRBs taken from a GRB pulse sample \citep{2021ApJ...919...37H}. Duration of the second pulse $\tau_2$ is shown relative to the separation between pulses ($sep$), where both timescales have been corrected for cosmological redshift by normalizing to the duration of the first pulse ($\tau_1$). Pulse durations increase as the separation between pulses increases. This increase is consistent with lateral jet motion which is either slowing or in which the jet profile is dispersing.
\label{fig:multi-pulse}}
\end{center}
\end{figure}

Figure \ref{fig:multi-pulse} demonstrates that pulse durations are not constant. In fact, they increase as pulse separations increase. This behavior is not consistent with periodic motion of a ballistic jet nozzle.  
Because pulse durations increase as separations increase, and because this increase is accompanied by an apparent decrease in pulse intensity and energy, there is a systematic evolution of the system. This evolution appears to be consistent with expansion of the emission surface described in Section \ref{sec:firehose}, as ejecta released each time the jet recrosses the line-of-sight will have to travel farther to reach the expanding emission surface.

\section{Discussion} \label{sec:disc}

GRB models have been around for over fifty years. In that time, the database of GRB behaviors has grown even as the mechanism responsible for producing their emission has remained elusive. Given the many models that have been considered over time, the idea that GRBs might be caused by moving jets is not a new one. For example, \cite{1996ApJ...473L..79B} considered the possibility that GRBs might result from ``cosmological relativistic blob emitting neutron star jets that precess past the line of sight." \cite{1995Ap&SS.231..191F} hypothesized that GRBs resulted from precessing neutron star jets (NSJ) in binary systems with solar-like companions. \cite{1999A&AS..138..503P} developed a model in which a precessing luminosity cone in a slaved accretion disk could be responsible for ``complex temporal behavior of bright gamma-ray bursts." \cite{2023ApJ...945...17G} have considered that jet precession may be a better explanation for GRBs with repeating emission episodes than millilensing, but this model also hypothesized that the mechanism responsible for the emission originates in radial interactions of material in the jet. Fargion and various co-authors ({\em e.g.,} \cite{1996AIPC..384..749F, 2003ChJAS...3..472F, 2006ChJAS...6a.342F}) have explored GRB light curves as being due to complex, short-term precession of narrow jets overlaid on longer timescale precessional motions. These motions have been designed to account for the shortest pulse structures, but not for the more recent discover of time-reversed GRB profiles. Recently, \cite{2021ApJ...916...71H} have considered that jet precession may contribute energy injections to GRB afterglow light curves.
However, these models were not designed to account for what appears to be the most direct evidence of a laterally-moving jet: the time-reversed structure of GRB pulses.

The time-reversed structure observed in $86\%$ of GRB pulses, as well as the general morphologies of GRB pulse types, can be directly explained by lateral jet motion across the observer's line-of-sight. 
For time-reversed structure to be observed, the jet itself must produce a nonuniform brightness distribution that is characterized by some sort of jet axial symmetry (see Section \ref{sec:time-rev}). On top of this, both the rigid/semi-rigid jet model and ballistic jet model described here require the jet nozzle to move laterally. The mechanism for this motion is not specified here, but may possibly be due to the instability of the accretion disk around the forming black hole.

Processes capable of causing relativistic jets to form and launch are associated with energy extraction from rotating compact objects. The most commonly accepted of these is the Blandford-Znajek process in which jets are powered electromagnetically by the winding of magnetic field lines interlacing a rotating central compact object \citep{1977MNRAS.179..433B, 2001MNRAS.326L..41K, 2010ApJ...711...50T}. By this process, a jet forms via conversion of the compact object's rotational energy into a Poynting flux-dominated outflow. For ultra-relativistic jets like those found in GRBs, this process is accompanied by instabilities so strong that they can cause the jet to kink \citep{2016MNRAS.456.1739B}, even preventing it from penetrating the star \citep{2014ApJ...785L..29M}. From a na{\"\i}ve perspective, the total rest mass available to a GRB system having 100 solar masses or less is around $10^{56}$ ergs. If we consider GRB energies in the range $10^{48}$ ergs to $10^{52}$ ergs, then $10^{56}/10^{48} =10^8$. Thus, the ratio of the (energy available for a magnetic field)/(jet power), where the magnetic field is providing containment, would be less than what is expected for conventional gamma-ray pulsars. As a result, the ability of the field to tightly constrain the jet direction would be considerably diminished, and rigid/semi-rigid moving jets seem less likely than jets with moving nozzles.

How lateral jet motion produces pulse asymmetry and spectral evolution is model-specific. The emission surfaces of rigid/semi-rigid jets must cross the observer's line-of-sight within a limited range of distances from the jet nozzle, and these distances are too close to the progenitor for generally-accepted astrophysical mechanisms to work. Such jets naturally develop large deviations needed to produce the pulse asymmetries and stretching factors observed in many GRB pulses. Contrarily, ballistic jets can easily produce photons at distances consistent with standard jet emission mechanisms because constraints on the distance of the jet emitting region from the nozzle are not present. Some amount of pulse asymmetry can occur naturally in the ballistic jet model while the jet emitting region is pushed outward, and additional asymmetry can be easily created as the nozzle's lateral motion slows and/or as the jet opening angle widens.

As jet ejecta loses energy in the surrounding medium, the jet volume increases and the energy density decreases, and the jet containment weakens. These effects should cause the emission spectrum to gradually soften, as expected for the hard-to-soft spectral evolution observed in both GRB pulses and GRBs overall.

Multi-pulsed GRBs are uncommon, but when they are observed they are consistent with a process that weakens over time. In the case of a rigid or semi-rigid jet, subsequent pulses might indicate that an existing jet is recrossing the line-of-sight, or perhaps they might indicate that additional jets have been launched from the nozzle. In the case of a ballistic jet, subsequent pulses might indicate that the nozzle has reactivated, or perhaps moved back in the direction of the observer. The increase of pulse durations with time is consistent with emission surface expansion in ballistic jets.

\section{Conclusions and Future Work} \label{sec:conc}

Laterally-moving jets, such as those described here, can explain a wide range of previously-unexplained temporal characteristics of GRB light curves. These include the time-reversed structures found in GRB pulses, asymmetric pulses and their structures, the rarity of GRB pulses, and correlations between increasing pulse durations and pulse separations. These characteristics are not addressed in most GRB models.
Furthermore, the relativistic properties of a jet as it crosses the line-of-sight are directly tied into pulse brightness, spectral hardness, and duration. GRB pulses may turn out to be an exceptional laboratory for studying the effects of special relativity. 

Lateral motion of the jet nozzle is more likely to occur as a jet is forming, and when the structure of the progenitor is relatively unstable. Misalignment of the rotation axis and magnetic field axis can occur during compact object formation, especially in the case of a tilted accretion disk. This process can produce a nozzle that moves in a chaotic fashion. However, a more stable configuration is also possible where nozzle precesses. In this case, the inner jet regions can spiral under the outer layers and turn the jet into an expanding spiral or helix. 

Scenarios by which relativistic jet curvature can develop have been examined by various authors using 2D and 3D magnetohydrodynamic (MHD) computer codes \citep{1993MNRAS.260..163R, 1996MNRAS.282.1114C, 1998A&A...334..750F, 2002ApJ...568..733M, 2015A&A...574A.143M, 2022ApJ...933...71F}. The two most common scenarios by which the moving jet might be produced are a) via a precessing jet nozzle arising from accretion into a compact source (similar to the massive star core-collapse model used to describe long-duration GRBs), and b) by orbital motion of material in a close binary system (similar to the merging neutron star model used to describe short-duration GRBs). Despite being launched from progenitors similar to those expected for GRBs, the jets produced in these models curve at large distances from the nozzle and move at slow enough velocities that the crossing time should be many orders of magnitude larger than the durations of GRB pulses. Thus, both the mechanism needed to produce a rigid or semi-rigid jet and the mechanism needed to produce photons from this jet are inconsistent with the close distances at which a rigid/semi-rigid jet would need to cross the line-of-sight in order to produce a GRB pulse. It is much more straightforward if the jet's nozzle moves across the line-of-sight and the jet simply sprays some of the ejecta in the direction of the observer. Gamma-rays can be produced by any of a number of standard processes when the ejecta is far removed from the source. Jets produced in the manner will be curved, but observers only see emission when the jet points at them. 

It has been previously demonstrated, upon accounting for time-reversed structure, that there are a relatively small number of GRB pulse light curve morphologies. This argues that there is some mechanism common to all GRB pulses. We propose that this commonality may be the jet structure itself, in which the smooth pulse component describes the core while the structured component describes the variations across the face of the jet. Each GRB pulse light curve allows the observed to view a slice of the jet surface. We have demonstrated here that many variations in GRB pulse light curves may simply result from jets passing off-axis across the observer's line-of-sight.

Two models have been proposed here to describe a laterally-moving jet. The rigid/semi-rigid jet model is able to satisfy many of the geometric concerns imposed by the observations, but the mechanism needed to produce gamma-ray emission must occur so close to the progenitor that it contradicts what is understood in standard GRB emission models. The ballistic jet model satisfies most geometric concerns by simply spraying relativistic material towards the observer as it crosses the line-of-sight, and using standard arguments that the jet ejecta cannot produce radiation until far from the progenitor, standard jet emission mechanisms remain viable. In both of these models, observers briefly see jet structure as the photon-producing regions of the jet cross the line-of-sight.  

The models developed here suggest that the number of pulses within a burst is an indicator of the level of activity in the jet's lateral motion. The time-reversed structure found in GRB pulses indicates that there must be only a few line-of-sight jet crossings because the number of pulses per GRB is observed to be small (typically one or two per burst). This implies either that GRB jets are relatively stationary or they are active but only rarely point in the observer's direction. for either of these options, the temporal spacing of GRB pulses supports the idea that jets lose energy over time. Two other pulse spacing issues provide additional clues: 1) late flares observed during GRB afterglows suggest that activity continues long after the main bursting event has been completed: either the jet or the nozzle are still undergoing lateral motion at late times, and energy is still available to power the nozzle, and 2) some GRB pulses overlap, suggesting that the nozzle or jet can undergo rapid fluctuations in structure and/or motion. 

Laterally-moving relativistic jets are able to account for a wide range of previously-unexplained GRB pulse behaviors such as those described in this paper. Despite the model's success, there are still unanswered questions. For example, we have used the pulse structure's stretching factor $s$ as an alias for the pulse asymmetry $\kappa$, when these values do not in fact represent a one-to-one correlation. Rather, the $s$ varies relative to $\kappa$ in such a way to somehow produce different pulse morphologies. Furthermore, some pulses (primarily symmetric crown pulses and u-pulses) have $s$ values that exceed unity, such that the structure is compressed on the decay side of the pulse relative to the rise side. This is contrary to the time asymmetry seen in the underlying pulse, which, when present, generally undergoes a rapid rise and a long decay. We are currently undertaking more detailed analyses to explore how models can explain these important second-order effects.

\begin{acknowledgments}
We are grateful to the feedback of several anonymous referees who greatly helped improved this paper. We also thank Charles A. Meegan, Amy Y. Lien, Istvan Horvath, and Peter Veres for helpful discussions.
\end{acknowledgments}

%




\bibliography{Hakkila}{}
\bibliographystyle{aasjournal}



\end{document}